\documentclass{aa}
\usepackage{graphics,rotate,psfig}
\def\sbu{${\rm mag\,\,arcsec^{-2 }}$\ }
\def\xbu{${\rm counts\,\,ksec^{-1}\,\,arcmin^{-2}}$\ }
\def\ergsec{${\rm erg\,\,s^{-1}}$}
\def\uflux{${\rm erg\,s^{-1}\,cm^{-2}}$}
\def\hi{H\,{\small I}}
\def\msun{${\rm M_{\odot }}$}
\newcommand \ra[4]{{#1}$^{\rm h}${#2}$^{\rm m}${#3}$^{\rm s}\!$.{#4}}
\newcommand \dec[3]{{#1}$^{\circ }\!\!$ {#2}\arcmin\,{#3}\arcsec}
\def\pano{\rule[0.0ex]{0cm}{2.5ex}}
\def\kato{\rule[-1.25ex]{0cm}{1.25ex}}
\newcommand{\mima}[2]{$_{\scriptscriptstyle -#1}^{\scriptscriptstyle +#2}$}
\def\countsec{ ${\rm counts\,\,sec^{-1}}$ }
\def\ha{H$\alpha$\  }
\def\kmsec{km\,\,s$^{-1}$\ }
\begin{document}

   \thesaurus{06  
              (11.09.1 NGC\ 7714;  
               11.09.1 NGC\ 7715;  
               11.09.2;  
               11.19.3;  
               11.16.1;  
               13.25.2)}  
   \title{X-ray observations of the interacting system Arp\ 284}

   \author{P.\ Papaderos\thanks{Visiting Astronomer, German-Spanish Astronomical Centre, 
Calar Alto, operated by the Max-Planck-Institute for Astronomy, Heidelberg,
jointly with the Spanish National Commission for Astronomy}
 \and K.J.\ Fricke}

   \offprints{P.\ Papaderos}

   \institute{Universit\"ats Sternwarte, Geismarlandstra\ss e 11, 
             37083 G\"ottingen, Germany}

   \date{Received \dots ; accepted \dots }

   \titlerunning{X-ray observations of the interacting system Arp\ 284}
   \authorrunning{P.\ Papaderos \& K.J.\ Fricke}

   \maketitle

   \begin{abstract}
We discuss the X-ray properties of the interacting system Arp\ 284, consisting of the 
active nuclear starburst galaxy NGC\ 7714 and its post-starburst companion NGC\ 7715.
A morphological signature of the interaction, thought to have started $<100$ Myr ago, 
is an asymmetric stellar ring dominating the intensity profile of NGC\ 7714 in the 
inner disk ($\sim 2$ exponential scale lengths).
In agreement to previous \emph{Einstein}-data our ROSAT~PSPC exposure shows the X-ray 
emission of Arp\,284 to be confined to NGC\ 7714. 
The bulk of the intrinsic X-ray luminosity in the ROSAT 0.1--2.4 keV band can be
accounted for by thermal emission from hot ($\sim 5\times 10^6$ K) gas and amounts to   
$\sim 2-4\times 10^{41}$ \ergsec.
Follow-up observations with the ROSAT~HRI revealed two distinct extended emitting regions 
contributing to the X-ray luminosity in NGC\ 7714. 
The more luminous of them ($L_{\rm X}\la 2\times 10^{41}$ \ergsec) coincides with the 
central starburst nucleus and can be explained this way. 
The fainter one ($L_{\rm X}\sim 8\times 10^{40}$ \ergsec)  
is located $\sim 20$\arcsec\ off-center and does not have any conspicuous optical counterpart. 
It is, likely, located at the borderline between the stellar ring and 
a massive ($>10^9$ \msun) \hi-bridge further to the east possibly intersecting NGC\ 7714.
The \hi\ and X-ray morphology and the extensive starburst nature of the 
nuclear energy source suggest different scenarios for the formation of the eastern 
emission spot. The possibilities of (i) collisional heating of the outlying gas by 
a starburst-driven nuclear wind and (ii) infall of \hi-clouds from the bridge onto the disk 
of NGC\ 7714 are discussed.
   \end{abstract}
\section{Introduction} 
Observational work during the past two decades has established the paradigm of interaction-induced 
activity in galaxies (cf. Fricke \& Kollatschny 1989; Sulentic 1990). 
Galaxy interactions were initially studied theoretically by simplified N-body simulations 
(Toomre \& Toomre 1972) and investigated in more detail by N-body/SPH models 
(Barnes \& Hernquist 1996 and references therein). Interactions are linked to a variety 
of physical processes such as nuclear starbursts, non-thermal nuclear activity, 
and morphological evolution into different classes of galaxies (Barnes \& Hernquist 1992). 
Already, collisionless calculations can simulate the transformation of merged 
spirals into ellipticals (Schweizer 1982) 
and the formation of tidal dwarf galaxies  (Duc \& Mirabel 1994; Sanders \& Mirabel 1996).
Thereby, understanding galaxy interactions is a prerequisite
for addressing issues like the morphological mix and the 
luminosity and number distribution of galaxies.
Gravitational perturbations of the gas component\linebreak through encounters between spirals 
can provoke large-scale internal gas instabilities, loss of angular momentum, and inflow of gas 
into the nuclear regions. 
Accumulation of gas at high densities (Solomon et al. 1992) is thought to be a necessary condition
for the onset of vigorous nuclear starbursts (Sanders \& Mirabel 1996 and references therein) 
or even the development of AGN activity. In the presence of dust,
the radiative output of massive stellar clusters is absorbed and reradiated in the 
far-infrared, a process which is thought to operate in the class of FIR-luminous galaxies. 
At the same time, the collective output of energy and momentum from OB stars and supernovae explosions
into the interstellar medium creates a hot ($\sim $ few $10^6$ K) X--ray emitting gas phase.
The thermal pressure exerted by this component on the ambient cold gas may terminate the gas-inflow 
and even disrupt the outlying \hi-layer, thus giving rise to a fast ($\sim 10^3$ km\ sec$^{-1}$) galactic wind 
(Heckman et al. 1987;1996, Veilleux et al. 1994). 
\begin{figure*}
\rule{0pt}{7cm}
\vspace*{-2mm}
{\caption[]{Contour map of Arp\ 284 in the R band. Contours, corrected for galactic 
extinction (0.085~mag), are shown between levels 16.3 \sbu to 
24.3 \sbu in steps of 0.5 mag. The nuclear regions of both galaxies are separated by 
1\farcm 96 (22.8 kpc). The marked stellar ring in NGC\ 7714 is best visible 
at $\sim 21.25$ \sbu. 
The position of the background QSO {[}HB89{]} 2333+019 at $z$: 1.871 is indicated. 
Note the faint ($\mu _{\rm R}\ga 23$ \sbu ) optical bridge connecting both systems. 
The insets to the lower-left and upper-right display the central 
20\arcsec $\times $ 20\arcsec and 50\arcsec$\times $50\arcsec\ 
regions of NGC\ 7715 and NGC\ 7714, respectively, after processing 
with the hb-method.}}
\label{fig1}
\end{figure*}

On the other hand, interactions between galaxies may effect a redistribution of 
a large fraction of the cold gaseous component on scales of hundreds of kpc. 
\hi-plumes or bridges emanating from the nuclear regions or protruding from the plane do 
not seem to be uncommon among interacting/merging galaxies. 
Gaseous features on scales of 50 to 150 kpc and mass ranges between 1.5 and 
$\sim 6\times 10^9$ \msun\ were discovered in systems spanning a wide range across the merging
sequence, such as Arp\ 144 (Higdon 1988), 
Arp\ 215 (Smith 1991), NGC\ 520 (Norman et al. 1996) and NGC\ 3256 (Hibbard \& van Gorkom 1996).
The fate of this material, in particular the question, whether it is going to be recaptured 
in the gravitational field of the merged system, is uncertain. 
It is, however, conceivable that after some time reinfall of tidally ejected matter 
onto a merger (Hibbard \& Mihos 1995) could lead to shock formation, thereby provoking extranuclear activity 
or feeding a subsequent burst of star formation.\\
%
\begin{figure}
\rule{0pt}{6cm}
\caption[]{Surface brightness profile of NGC\ 7714 in the R-band. The profile is not corrected for inclination. 
The luminosity distribution of the disk component (solid light line) is obtained by a linear fit to 
the outer part ($R^*>25$\arcsec) of the profile, weighted by the photometric uncertainties. 
Open symbols display the residual intensity distribution in excess of the 
disk component.
The intensity decrease by roughly 0.4~mag in the surface brightness profile of 
NGC\ 7714 at $R^*\approx 20$\arcsec\ reflects the transition from the 
(bar$+$ring) component to the disk-dominated regime.
The angular distance below which the disk dominates the light is 25\arcsec\ 
corresponding to $\sim 2$ disk exponential scale lengths (2.39$\pm$0.6 kpc).
The luminosity distribution of the compact starburst nucleus is shown by the 
thick solid line. Its apparent luminosity integrated for $\mu _R\leq 25$\ \sbu 
is determined to 13.5~mag, equivalent to $\sim 50$\% of the 
luminosity of NGC\ 7714 in excess of the disk.}
\label{fig2}
\end{figure}
%
In this paper we focus our attention on the X-ray properties of the 
interacting system Arp\ 284 (Arp 1966). This system consists of 
the nuclear starburst galaxy NGC\ 7714 and its fainter inactive companion NGC\ 7715. 
Weedman et al. (1981) have first pointed out the extraordinary properties of NGC\ 7714 among 
starburst nuclei and inferred from \emph{Einstein} observations 
an intrinsic X-ray luminosity of $0.6\div1\times 10^{41}$ \ergsec\ 
in the 0.25--3.5 keV energy range. 
Other observational support for the presence of an ongoing burst of star formation in 
NGC\ 7714 is provided by the detection of WR spectral features and strong H$\alpha $ 
emission on scales of $\sim 12$\arcsec\ (Gonz\'alez-Delgado et al. 1995, 
Smith et al. 1997, Telles \& Terlevich 1997). 
Spectral synthesis models of Bernl\"ohr (1993) imply for NGC\ 7714 a burst age of $\sim 20$ Myr. 
The same author concluded that the less massive interaction counterpart NGC\ 7715 
has undergone a burst of star formation roughly 70~Myr ago (Bernl\"ohr 1993), 
presumably at the time of the initial impact between both galaxies (Smith et al. 1992). 
Arp\ 284 displays features predicted to evolve shortly after the first 
closest passage between two gas-rich spirals (Smith et al. 1997, Hernquist 1992). 
A striking one is a large asymmetric ring located $\sim 20$\arcsec\ eastwards of 
the nucleus of NGC\ 7714.
Narrow-band imaging of NGC\ 7714 has not revealed a notable contribution of 
H$\alpha $ emission to the light of this feature (Gonz\'alez-Delgado et al. 1995, 
Smith et al. 1997). Also high angular resolution 21\,cm VLA-maps of Arp\ 284 
(Smith et al. 1997) show that the ring is comparatively devoid of \hi-gas.
Furthermore, Bushouse \& Werner (1990) have reported that the stellar ring 
possesses NIR colours similar to those in the disk. All latter facts 
lend support to the idea that its formation process is related to a collisionally 
induced dynamical perturbation of the stellar disk (Lynds \& Toomre 1976) rather than being
the result of past star formation. This hypothesis is in accord with numerical simulations of 
off-center encounters between spiral galaxies demonstrating the formation of similar 
morphological features (Mihos \& Hernquist 1996, Smith et al. 1992,1997).
In the following we investigate the X-ray properties of Arp 284 using recent ROSAT data, together 
with relevant optical and radio observations.
We shall adopt a distance of 40~Mpc (H$_0$=75 km\ sec$^{-1}$ Mpc$^{-1}$) to Arp\ 284 (Bernl\"ohr 1993).
\section{Optical data}
Arp\ 284 was observed with the 2.2m telescope at Calar Alto with a SITe-CCD detector attached to 
the CAFOS focal reducer giving an instrumental scale of 0\farcs 53\,pixel$^{-1}$. 
The target was observed for 3 min in the Johnson R under non-photometric conditions
with a seeing of 2\farcs 1$\pm$0\farcs 05.
The image was calibrated on the basis of aperture photometry by Bushouse \& Werner (1990) and Bernl\"ohr (1993).
The detection limit of point sources at a S/N of 2 was estimated to $\sim 20.3$ R\,mag.
As shown in Fig.\,1, the nuclear regions of the interacting galaxies NGC\ 7714 and NGC\ 7715 are separated by 
$\sim 1\farcm 96$, corresponding to a linear distance of 22.8 kpc. 
Both galaxies are apparently connected by a diffuse and faint ($\mu _{\rm R}\ga 23$ \sbu) stellar bridge protruding 
from NGC\ 7715 into the outer boundary of the stellar ring in NGC\ 7714. 
The point source indicated to the upper-left is the 
QSO {[}HB89{]}\,2333+019 ($\alpha,\delta$)$_{2000}$: \ra{23}{36}{30}{5},\dec{+02}{10}{42} (Hewitt \& Burbidge 1989). 
The apparent luminosity of this background source ($z=1.871$) was determined to $m_{\rm R}=18.6\pm 0.1$~mag.

For the sake of revealing faint compact structures we processed the central region of NGC\ 7715 and 
NGC\ 7714 with a hierarchical binning algorithm (hereafter hb-method) designed for extracting 
weak features of angular size smaller than a user-defined angular threshold.
Results obtained by the hb-method for the central regions of NGC\ 7714 and NGC\ 7715 are displayed in 
the insets in Fig.\,1. The dominant features in NGC\ 7714 within a region of $\la 12$\arcsec\ in diameter are
the compact starburst nucleus and an adjacent fainter knot. As shown in Fig.\,2
the intensity distribution of NGC\ 7714 can be approximated reasonably well by an exponential fitting 
law for $\mu _R\ga 22.5$ \sbu. 
The marked intensity depression by $\sim 0.4$ mag in the SBP of NGC\ 7714 at a photometric radius $R^*\sim 20$\arcsec\  
reflects the sharp transition from the (bar$+$ring)--dominated regime of NGC\ 7714 to its fainter disk population.
%
\begin{figure}
\rule{0pt}{9cm}
\caption[]{ROSAT~PSPC contours of the 0.1--2.4 keV emission of Arp\ 284 superimposed on an optical image. 
The X--ray image was convolved with a Gaussian with fwhm 35\arcsec, equivalent to the average resolution 
of the PSPC camera in the 0.1--2.4 keV energy band. The positions of the QSO {[}HB89{]} 2223+019 and of the 
starburst nucleus of NGC\ 7714 are indicated by crosses. The X--ray source designated \#1
is presumably associated with a faint background galaxy (cross). 
Contours correspond to intensities of 0.75,\,1,\,2,\,4,\,8,\,12,\,16 and 22 \xbu \emph{above} 
the background (0.87 $\pm$ 0.36 \xbu).}
\label{fig3}
\end{figure}
\section{X-ray data}
We observed Arp\ 284 in the 0.1--2.4 keV energy band with both, the Position Sensitive Proportional 
Counter (PSPC) and the High Resolution Imager (HRI) on board ROSAT 
(Tr\"umper 1983, Pfeffermann et al. 1987). The analysis of the X--ray data was carried out with the 
EXSAS package (Zimmermann et al. 1994) implemented in the MIDAS.
\subsection{PSPC observations}
Arp\ 284 was observed with the PSPC twice (PSPC\_1: 1.33 and PSPC\_2: 12.68 ksec). 
%
\begin{figure*}
\begin{picture}(0,6)
\rule{0pt}{6cm}
%
\end{picture}
\caption[]{\emph{(left)} ROSAT~PSPC spectrum of NGC\ 7714.
The dotted curve displays the channel pulse height distribution of the 
X-ray background. The hardness ratios of the spectrum, $HR_1=0.87\pm0.03$ and $HR_2=0.20\pm0.13$ suggest
a highly absorbed source with an intermediately hard intrinsic photon distribution.
\emph{(middle)} Unconstrained thermal bremstrahlung \emph{(thbr)} fit to the soft X-ray 
spectrum of NGC\ 7714. Residuals to the observed flux distribution are displayed at the 
bottom panel. \emph{(right)} $\chi ^2$ map for a \emph{thbr} model
as function of the absorbing gas column density (N$_{\rm H}$) and the mean plasma temperature ($k$T). 
The contours correspond to confidence levels of 68.3, 95.4 and 99.7\%. The deduced fit solution (Table\,1) 
is indicated by the cross.}
\label{fig4}
\end{figure*}
\begin{table*}
\caption[]{Model fits to the ROSAT PSPC spectrum}
\begin{tabular}{lccccccc}
\hline
Model  &  {\rm N}$_{\rm H}^{\mathrm{a}}$   & $k$T  & $\Gamma $  & Flux$^{\mathrm{b}}$ (0.1-2.4 keV)  &  Flux$^{\mathrm{c}}$ (0.1-2.4 keV) &
$L_{\rm X}$ (0.1-2.4 keV) & $\chi^2/d.o.f^{\mathrm{d}}$\pano \\
       &  10$^{21}$ cm$^{-2}$  & keV   &            & 10$^{-13}$ \uflux & 10$^{-12}$ \uflux & 10$^{41}$ \ergsec         &        \\
\hline
{\it thbr}   & 1.7\mima{0.99}{2.56} & 0.55\mima{0.25}{0.46}               & ---      
 & 2.39\mima{0.84}{3.31}  & 1.04\mima{0.36}{1.43} & 2.0\mima{0.70}{2.8}   & 0.51\pano  \\
{\it thbr}{\scriptsize (F1)}   & 2.2  & 0.47\mima{0.03}{0.08}               & ---      
 & 2.28\mima{0.03}{0.04}  & 1.41\mima{0.02}{0.03} & 2.7\mima{0.04}{0.06}   & 0.45  \\
{\it thbr}{\scriptsize (F2)}   & 2.55 & 0.43\mima{0.03}{0.08}               & ---      
& 2.22\mima{0.03}{0.04}  & 1.74\mima{0.02}{0.03} & 3.3\mima{0.04}{0.06}   & 0.48  \\
{\it powl}   & 0.9\mima{0.2}{0.6} & ---  &  2.3   & 2.80\mima{0.49}{0.32} & 
0.83\mima{0.15}{0.10} & $\!\!$1.6\mima{0.3}{0.2}  &   1.19  \\
{\it bbdy}   & 0.5\mima{0.34}{1.97} & 0.24\mima{0.05}{0.04} & --- & $\!\!$2.30\mima{1.5}{0.9} & 0.31\mima{0.20}{0.12} &
0.6\mima{0.39}{0.23} & 0.52\kato \\
\hline
\end{tabular}

\vspace*{1ex}
$^{\mathrm{a}}$ Absorbing gas column density from a foreground screen model combining the
Galactic \hi\ column density $n_{\rm H}^{\rm G}=0.493\times 10^{21}$ cm$^{-2}$ 
to the direction of Arp~284 (Dickey \& Lockman 1990) with
the intrinsic column density $n^{\rm i}_{\rm H}$ of the target.\\
\emph{thbr}: Unconstrained fit. N$_{\rm H}$ has been fitted from the X-ray data.\\
F1,F2: Constrained fits. The average intrinsic absorption $n_{\rm H}^{\rm i}$ is estimated from the \hi\ VLA-maps 
(Smith et al. 1997) on the basis of two different assumptions. 
F1: Each of the X-ray emitting regions within NGC\ 7714 (cf. Sect.\,3.2) experiences a different 
amount of intrinsic absorption, and
F2: Either emitting source resides midway in the \hi-disk of NGC\ 7714 and experiences 
the same amount of intrinsic absorption.\\
$^{\mathrm{b}}$ \emph{absorbed} flux.\\
$^{\mathrm{c}}$ \emph{intrinsic} flux.\\
$^{\mathrm{d}}$ $\chi ^2$ per degree of freedom (\emph{d.o.f.})
\end{table*}
In both observations, intervals with a count rate exceeding by 3$\sigma $ the mean value were screened out. 
We also checked that the aspect solution provided by the Standard Analysis Software System (SASS, Voges et al. 1992) does 
not contain residual aspect errors greater than 2\arcsec. 
After the latter corrections we obtained effective exposures of 1.3 ksec and 12.1 ksec 
for PSPC\_1 and PSPC\_2, respectively. 
Our ROSAT PSPC exposure (Fig.\,3) confirm the outcome of a preceding analysis of \emph{Einstein}-IPC data 
(Fabbiano et al. 1992), namely that the entire soft X-ray emission in Arp\ 284 is confined to NGC\ 7714. 
The source to the left was registered at a PSPC count rate of 
$(8.3\pm 2.8)\times 10^{-3}$ \countsec. 
Follow-up observations with the ROSAT HRI (Sect.\,3.2), show it
to coincide with the background QSO {[}HB89{]} 2333+019. 
Its PSPC-spectrum, extracted within a circular aperture of radius 100\arcsec, appears highly absorbed 
($HR_1$=+0.89$\pm $ 0.03)\footnote{Hardness ratios are obtained from the \emph{net} counts of a source as
$HR_1=(Hard-Soft)/(Hard+Soft)$ and $HR_2=(H2-H1)/(H1+H2)$ registered in the standard ROSAT PSPC bands 
{\it Soft} (0.1-0.4 keV), {\it Hard} (0.5-2.0 keV), {\it H1} (0.5-0.9 keV) and {\it H2} (0.9-2.0 keV).}
and intrinsically rather hard ($HR_2$=+0.39$\pm $0.17). 
The X-ray source designated \#1 is most probably associated to
a faint (m$_{\rm R}=17.2$) background galaxy at ($\alpha,\delta$)$_{2000}$: \ra{23}{36}{26}{6}, \dec{02}{13}{17}. 

NGC\ 7714 was detected at a PSPC count rate of (2.13 $\pm $ 0.45)$\times$10$^{-2}$ \countsec, yielding 
286$\pm $60 source counts for the coadded PSPC\_1 $+$ PSPC\_2 exposures.  
Its photon spectrum (Fig.\,4, left) was extracted from a circular aperture of radius 
135\arcsec\ and binned to a ${\rm S/N\sim 5.5}$. 
The count rates registered in the \emph{standard} ROSAT PSPC energy bands 
\emph{Soft}, \emph{Hard}, \emph{H1} and \emph{H2} are
$(0.13\pm 0.02)\times 10^{-2}$, $(1.9\pm 0.34)\times 10^{-2}$, 
$(0.74\pm 0.14)\times 10^{-2}$ and $(1.11\pm 0.21)\times 10^{-2}$ 
counts\,sec$^{-1}$, respectively. The corresponding hardness ratios, 
$HR_1=0.87\pm0.03$ and $HR_2=0.20\pm0.13$, suggest 
a highly absorbed, intermediately hard spectrum.
%
\begin{figure}[t]
\rule{0pt}{8.2cm}
%
\caption[]{Comparison of Arp\ 284 with a sample of interacting/merging galaxies (filled circles) 
compiled from Read \& Ponman (1998), Kollatschny et al. (1996), Wang et al. (1997), 
Fricke \& Papaderos (1996) and Papaderos \& Fricke (1998) in the 
log(L$_{\rm X}$) vs. log(L$_{\rm X}$/L$_{\rm B}$) (top) and 
log(L$_{\rm FIR}$/L$_{\rm B}$) vs. log(L$_{\rm X}$/L$_{\rm B}$) planes (bottom). 
The bars indicate the positional shift of Arp\ 284 in the diagrams when a range of the
intrinsic X-ray luminosity between 2 and $4\times 10^{41}$\ \ergsec\ is adopted. 
A linear fit (solid line, upper diagram) to the colliding galaxies implies a relation 
log(L$_{\rm X}$/L$_{\rm B}$)$\propto$ 3/4\ log(L$_{\rm X}$).
The mean ratio log(L$_{\rm X}$/L$_{\rm B}$) $=$ --3.83 for the sample of undisturbed spirals (open symbols; 
Read et al. 1997) is indicated by the dashed horizontal line (Sp).
The mean log(L$_{\rm X}$/L$_{\rm B}$) ratios for quiescent (N), starburst (SB), and Seyfert (Sey) galaxies 
investigated with \emph{Einstein}  (David et al. 1992;\,small dots) are shown by the thick lines.}
\label{fig5}
\end{figure}
%
The hardness of the observed photon distribution complicates a precise assessment  
of the intrinsic absorbing column density $n^{\rm i}_{\rm H}$ within NGC\ 7714. 
This is because $n^{\rm i}_{\rm H}$ when deduced from unconstrained spectral fits
is sensitively related to the spectral distribution below $\sim 0.7$ keV. 
Given that in the latter energy range the X-ray background becomes comparable or 
larger than the source flux the model-dependent $n^{\rm i}_{\rm H}$ can significantly 
vary depending on the background determination.
In order to study this effect we performed a number of tests 
by selecting background photons from different circular annuli and binning 
the resulting photon spectrum to a varying S/N. 
By applying a thermal bremsstrahlung model we have obtained 
two typical sets of solutions for $n^{\rm i}_{\rm H}$ and the plasma temperature $k$T 
differing only marginally in terms of $\chi^2$. 
The first one implies $n^{\rm i}_{\rm H};k{\rm T}$ 
$\sim $ $1.2\times 10^{21}$ cm$^{-2}$;$\la 0.55$ keV ($\chi^2/d.o.f\sim 0.5$;  
Table\,1) and the second one $n^{\rm i}_{\rm H},k{\rm T}$ $\sim $ 
$2.5\times 10^{21}$ cm$^{-2}$; 0.4 keV with ($\chi^2/d.o.f\sim 0.7$).
The corresponding lower and upper values for the intrinsic 0.1--2.4 keV luminosity of NGC\ 7714 are amounting to
$2$ and $4.4\times 10^{41}$ \ergsec, respectively. 
Next we shall investigate constrained fits, in which the intrinsic absorbing column density is 
estimated from HI-maps.
From VLA-studies, Smith et al. (1997) have determined within a region of 
11\arcsec$\times $8\arcsec\ centered at the starburst nucleus of NGC\ 7714 a mean \hi\ column 
density of $\sim 4.1\times 10^{21}$ cm$^{-2}$. As we shall show in Sect.\,3.2, the HRI-X-ray emission
of NGC\ 7714 is contributed by two distinct emitting sources. The most luminous one contributing
$\sim 2/3$ of the received photons coincides with the nuclear region while the second one 
is off-centered and coincides with a region with a lower \hi-column density. 
We may assume that the nuclear source resides midway in the \hi-disk of NGC\ 7714, thus 
experiencing an intrinsic absorption $n^{\rm i}_{\rm H}\sim 2.05\times 10^{21}$ cm$^{-2}$ and 
the extranuclear source $n^{\rm i}_{\rm H}\sim 1\times 10^{21}$ cm$^{-2}$. 
Weighting the latter values by the number of received photons we obtain an average
intrinsic column density $n^{\rm i}_{\rm H}$ $\sim 1.7\times 10^{21}$ cm$^{-2}$. 
A \emph{thbr} fit with ${\rm N_H}=n^{\rm i}_{\rm H}+ n^{\rm G}_{\rm H}$,
where  $n^{\rm H}_{\rm G}=0.49\times 10^{21}$ cm$^{-2}$ is the Galactic column 
density, results in an intrinsic luminosity of $2.7\times 10^{41}$ \ergsec\ (Table\,1;\,F1).
This luminosity estimate will increase to $L_{\rm X}\sim 3.3\times 10^{41}$ \ergsec, if one
assumes that either emitting region in NGC\ 7714 reside midway in the galactic plane, thus both
experience an intrinsic absorption $n^{\rm i}_{\rm H}$ $\sim 2\times 10^{21}$ cm$^{-2}$ (Table\,1;\,F2).\\ 
In view of the starburst nature of NGC\ 7714 we consider a thermal bremsstrahlung 
(\emph{thbr}) model to provide a reasonable approximation to its intrinsic 
soft X-ray spectrum.

A Raymond-Smith (1977) model was applied to the data assuming various elemental 
abundances, giving, however, unacceptable results in terms of $\chi ^2$ and 
implying in all cases an excessive absorbing gas column density ($>8\times 10^{21}$ cm$^{-2}$). 
A black-body fit to the spectrum leads to a $\chi^2$ value comparable 
to the one obtained by applying a \emph{thbr} model.
However, it implies no intrinsic absorption within NGC\ 7714 in disagreement 
to the radio measurements. 
A power-law \emph{(powl)} model, with a fixed photon index $\Gamma=2.3$,
which is the typical value found for extragalactic objects with ROSAT 
(Hasinger et al. 1991) yields an ${\rm N_H}$ $\sim 0.9\times 10^{21}$ cm$^{-2}$, 
most probably incompatible with the average gas surface density 
inferred by Smith et al. (1997). 
Moreover, given that optical spectroscopy 
does not indicate the presence of an additional AGN source in NGC\ 7714 we 
will discard the \emph{powl} model.

Adopting the mean value of $\sim 3\times 10^{41}$ \ergsec\ for the X-ray luminosity
and estimating the B and FIR luminosities from the RC3-- and the IRAS-PSC-catalogue 
following Tully (1988) and Devereux \& Eales (1989), 
respectively, we obtain the ratios log(L$_{\rm X}$/L$_{\rm B}$) $=$ --2.45 and 
log(L$_{\rm X}$/L$_{\rm FIR}$) $=$ --2.46 for Arp\ 284. 

In Figure\,5 we compare Arp\ 284 in the log(L$_{\rm X}$) vs. log(L$_{\rm X}$/L$_{\rm B}$) and
log(L$_{\rm FIR}$/L$_{\rm B}$) vs. log(L$_{\rm X}$/L$_{\rm B}$) diagrams 
with the location of the other interacting/merging galaxies: 
Arp\ 270, Arp\ 242, NGC\ 4038/9, NGC\ 520, Arp\ 220, NGC\ 2623, NGC\ 7252, AM\,1146-270 (Read \& Ponman 1998), 
Mkn\ 789, Mkn\ 1027 (Kollatschny et al. 1996), Mkn\ 266 (Wang et al. 1997),
NGC\ 6240 (Fricke \& Papaderos 1996) and Arp\ 278, Arp\ 215 (Papaderos \& Fricke 1998). 
Open circles show the sample of isolated spirals discussed in Read et al. (1997). 
Small dots show the sample of \emph{Einstein}-detected normal, starburst and Seyfert galaxies 
studied by David et al. (1992), adapted to H$_0$ $=$ 75 km\ sec$^{-1}$\ Mpc$^{-1}$.
Since the X-ray luminosities used by the latter authors refer to the 0.5--4.5 keV energy range
they were shifted by $\delta\log(L_{\rm X})=-0.045$ to adapt them to the ROSAT energy band. 
For this purpose we assumed a \emph{thbr} spectrum with $k$T $=$ 5\,keV for their sample.  
Arp\ 284 falls into the upper range in both diagrams for colliding galaxies.
Figure\,5a shows a narrow linear correlation between log(L$_{\rm X}$/L$_{\rm B}$) 
and log(L$_{\rm X}$): log(L$_{\rm X}$/L$_{\rm B}$) $\propto 0.75\pm 0.15\, \log({\rm L_X})$.
A similar but less pronounced trend is seen in Fig.\,5b correlating log(L$_{\rm X}$/L$_{\rm B}$) 
with log(L$_{\rm FIR}$/L$_{\rm B}$).
These trends are reminiscent of the correspondence between the flux ratios 
X/B with the flux ratio H$\alpha$/B of interacting galaxies (Wang et al. 1997) 
and support the view that the strong X-ray emission from 
colliding galaxies is not due to a global process but is closely tied 
to the starburst activity induced in such systems.
\subsection{HRI observations}
Arp\ 284 was observed with the HRI-camera for 45.86 ksec from 26 through 29 December 1994.
Each observation combines data recorded over several satellite orbital intervals (OBIs).
We verified that none of the OBIs was mispointed by comparing images constructed for each 
individual observation of duration greater than 1 ksec. After correction for residual aspect 
errors and rejection of time intervals with enhanced background emission we 
obtained a net exposure time of 45.6 ksec.
\begin{figure*}
\begin{picture}(16,10)
\rule{0pt}{10cm}
\end{picture}
{\caption[]{Contour map of the X-ray emission of NGC\ 7714 (\emph{thick lines}) 
as obtained from the ROSAT~HRI image, overlaid to a R-band image. The ROSAT\ HRI map was 
smoothed with a Gaussian with fwhm$=$5\arcsec. The optical image was processed by 
the hb-method (Sect.\,2) to better visualize fine features in the central region of NGC\ 7714.
X--ray contours correspond to intensity levels of 8,\,12,\,18,\,25,\,35,\,50,\,75 and 100 
\xbu \emph{above} the background (4.3$\pm $0.8 \xbu). The inset to the top-right shows the 
X--ray contours of the background QSO {[}HB89{]}\ 2333+019 at equal levels.
The optical extent of NGC\ 7714 at the extinction-corrected surface brightness 
of 23 R\ \sbu\ is shown by the outermost contour (\emph{thin solid line}). 
Thin light contours delineate the VLA-detected \hi-bridge eastwards of NGC\ 7714 
(Smith et al. 1997).
The X-ray emission of NGC\ 7714 is contributed by two distinct emitting regions 
with a count rate ratio $\sim 2$. 
The more luminous one (N) coincides with the nuclear starburst region while 
component E is located $\sim 20$\arcsec\ eastwards of the nucleus,
close to the outer boundary of the asymmetric stellar ring of NGC\ 7714 (Sect.\,2). 
Contrary to the inactive region subtended by the stellar ring, the nuclear region (N) as well 
as features {\bf\small I} and {\bf\small II} are found to be strong H$\alpha $ sources.}}
\label{fig6}
\end{figure*}
The astronomical position of the targets as provided by the SASS was checked 
by registering the position of the QSO {[}HB89{]} 2333+019 at the optical image (Fig.\,1). We deduced a residual 
aspect error of $\sim 8$\arcsec, compatible to the typical residual boresight error of 6\arcsec\ ascribed to SASS. 
The QSO was detected at a net source rate of ($1.9 \pm 0.7)\times 10^{-3}$\countsec, thereby making unfeasible to 
monitor residual wobbling errors in intervals shorter than the wobbling period (400 sec; cf. Bischoff et al. 1996) 
and improve on the short term aspect solution provided by the SASS. For the same reason other 
rectification methods, such as the one proposed by Morse (1994) and G\"udel \& K\"urster (1996) 
were not applied to the data.

Figure\ 6 displays the X-ray emission of NGC\ 7714 as obtained from ROSAT~HRI observations. 
The gray map shows a number of weak optical features in NGC\ 7714's central region, revealed 
by the hb-technique. The optical extent of NGC\ 7714 at an extinction-corrected 
surface brightness of $\mu _{\rm R}=23$ \sbu\ is shown by the thin contour. 
Optical features designated {\bf\small I} and {\bf\small II}, both located westwards of 
the nucleus, were found to possess H$\alpha $ emission 
(Bernl\"ohr 1993, Gonz\'alez-Delgado et al. 1995) implying an ongoing extranuclear star-formation activity.
The starburst nucleus indicated by the cross appears very compact (diameter $\la$ 7\arcsec) with a fainter 
knot located $\sim 5\farcs 7$ to the east. From this knot proceeds a faint arm connecting to region {\bf\small I}.
The overlayed thick lines are computed from the ROSAT~HRI exposure, convolved with a Gaussian 
with fwhm 5\arcsec, equivalent to the nominal on-flight resolution of the camera.
Figure\,6 reveals two discrete X-ray emitting components within the 
optical size of NGC\ 7714.
The most luminous one (N) coincides with the starburst nucleus of NGC\ 7714 
while component E is offset with respect to the former by $\sim 20$\arcsec. 
In view of possible interpretations of the origin of the extranuclear X-ray emitting component E
(Sect.\ 4.2) it is notable that its location does not coincide with any conspicuous compact optical feature 
but rather appears associated with the clumpy luminosity pattern defined by the asymmetric stellar ring. 
The number of \emph{net} counts in regions E and N are 112$\pm $16 and 221$\pm $22 (cf. Table\ 2), 
corresponding to count rates of 2.48$\pm $0.36 counts\,ksec$^{-1}$ and 4.85$\pm $0.46 counts\,ksec$^{-1}$, respectively.
Due to the low count rate a variability check for each knot separately proves not feasible. 
A variability check for NGC\ 7714 as whole carried out on both PSPC- and HRI-data has not 
revealed any systematic flux variations.\\ 
Unfortunately, the assessment of the spectral properties of each of the X-ray emitting 
components is being hampered by their low angular separation. 
While deconvolution of the PSPC data in the $H2$ band using an energy-weighted model for the
Point Response Function (PRF) has clearly demonstrated the E-W elongation of the source, the quality of the data 
does not permit to put firm constraints on the spatial variation of the PSPC hardness ratio 
(cf. Read et al. 1995).

The ROSAT HRI is known to possess some moderate energy resolution. Thus, we checked
the HRI softness ratios (counts in channels 1-5 divided by counts in channels 6-11; Wilson et al. 1992)
deduced to 1.1$\pm$0.2 and 1.6$\pm$0.5 for component N and E, respectively.
Although they suggest a slightly softer spectrum for component E they still, 
given the uncertainties, do not discard the possibility that either source share 
similar spectral properties. Furthermore, $\sim 90$\% of the flux of both components 
was received in channels 3-9 possibly making 
this method not appropriate for our data. 

Assuming that the spectral properties of either source can be approximated
by the unconstrained \emph{thbr} fit (Table\,1) one obtains for the HRI 
\emph{energy conversion factor} (ECF; RO\-SAT Technical Appendix) 
$6.24\times 10^{-2}$.
Thereby, the source count rates translate to intrinsic fluxes of $(3.97 \pm 0.6)\times 10^{-13}$ \uflux\  and 
$(7.78\pm 0.75)\times 10^{-13}$ erg s$^{-1}$ cm$^{-2}$ for regions E and N, respectively,
corresponding to component luminosities of $(7.6\pm 1.1)\times 10^{40}$ \ergsec\ and 
$(1.49\pm 0.15)\times 10^{41}$ \ergsec.\\ 
From case F1 in Table\,1 (e.g. each source experiences a different intrinsic absorption)
one obtains ECFs of $4.225\times 10^{-2}$ for the N and $6.0\times 10^{-2}$ for the E sources.
The corresponding intrinsic fluxes and luminosities are then 
$f^{\rm N}=(11.5\pm 1.1)\times 10^{-13}$ and $f^{\rm E}=(4.13\pm 0.6)\times 10^{-13}$ \uflux\ e.g. 
$(2.2\pm 0.2)\times 10^{41}$ and $(7.9\pm 1.1)\times 10^{40}$ \ergsec\ for N and E, respectively.\\
The summed luminosity of $\sim 3\times 10^{41}$ \ergsec\ is again compatible to the 
$L_{\rm X}\sim 2.7\times 10^{41}$ \ergsec\ obtained from the fit.
In summary, both estimates yield consistently for component E an X-ray luminosity of 
$\sim 8\times 10^{40}$ \ergsec\ and a luminosity ranging between 1.5 
and $2.2\times 10^{41}$ \ergsec\ for the nuclear source. 
%
\subsection{Extent of the components} 
%
The assessment of the true extent of compact sources revealed by the HRI
may be complicated by the fact that, in an individual pointing, the theoretical 
resolution of the ROSAT~HRI camera ($\sim 5$\arcsec) may be degraded by residual 
errors in the correction for the periodic wobbling of the satellite. 
Such errors may cause elongations in point sources on scales of $\la 10$\arcsec\ (cf. David et al. 1993) 
and lead to inflated radial intensity profiles.

Figure\,7 shows the background-subtracted intensity profiles of both X-ray components in NGC\ 7714 as 
derived from the smoothed ROSAT~HRI map (Fig.\,6). The total intensity profile of NGC\ 7714, containing 
some additional faint emission at levels below $\sim 10$ \xbu is shown by filled squares. 
The thick curve shows an empirical approximation to the on-flight PRF of the ROSAT~HRI 
(David et al. 1993) convolved with the same smoothing function as the one applied to the 
X--ray map (Fig.\,6) and scaled to the maximum intensity of component E. 
Numerical integration of the smoothed PRF yields for its effective radius $r_{50}=3\farcs 3$ and 
for the radius containing 80\% of the flux $r_{80}=5\farcs 9$. 
The corresponding quantities for each X-ray component, as obtained from profile integration to a 
threshold of 1 count ksec$^{-1}$ arcmin$^{-2}$, are $r_{50}=(6\farcs 4\pm 0\farcs 2)$ and 
$r_{80}=(10\farcs 9\pm 0\farcs 25)$ for the nuclear component N 
and $r_{50}=(5\farcs 6\pm 0\farcs 2)$ and $r_{80}=(9\farcs 0\pm 0\farcs 3)$ for component E, 
thereby incompatible to those expected for a point source.\\
This result requires, however, a further check since,
as pointed out above, the actual PRF of 
an individual HRI-pointing may differ from the nominal one.\\
For this purpose we assumed that the true PRF at the location of NGC\ 7714 can 
be approximated by the intensity distribution of the background QSO {[}HB89{]} 2333+019 
(Fig.\,7, thin line), located $\approx 3$\arcmin\ off-axis the RO\-SAT HRI field.
For the latter source we obtain $r_{50}=4\farcs 8$ and $r_{80}=7\farcs 7$, 
e.g. values by $\sim 40$\% greater than those corresponding to the nominal PRF. 
The magnified PRF is presumably attributable to residual errors in the 
satellite's aspect solution rather than to the slightly degraded PRF 
at the off-axis angle of {[}HB89{]} 2333+019.

Nevertheless, the latter comparison demonstrates that the characteristic radii 
$r_{50}$ and $r_{80}$ for either X-ray emitting region in NGC\ 7714
are larger than both, the nominal and actual PRF of the ROSAT~HRI exposure. 
Furthermore, it is relevant to the question of the extent of these components
that, for intensities fainter than 30 counts ksec$^{-1}$ arcmin$^{-2}$, 
their intensity distribution is systematically flatter than the one of the QSO. 
As a result, we conclude that the emission pattern of either X-ray emitting component
in NGC\ 7714 is definitely incompatible with that of a point source. 
\begin{figure}
\rule{0pt}{7cm}
\caption[]{Background subtracted intensity profiles of either X-ray emitting region revealed in NGC\ 7714 by 
the ROSAT~HRI (Fig.\,6). Both profiles were derived from the HRI-exposure after it was convolved with a Gaussian with 
fwhm$=$5\arcsec. The intensity profiles of the eastern (E) and the nuclear (N) component in units of \xbu\ are shown
with open and filled circles, respectively. The total emission at NGC\ 7714 is displayed by filled squares. 
The empirical approximation to the on-flight PRF of ROSAT~HRI, convolved with a Gaussian with fwhm$=$5\arcsec\ and 
scaled to the maximum intensity of component E, is shown by the thick curve.
The thin solid line shows the intensity profile of the background QSO {[}HB89{]} 2333+019 located 3\arcmin\ off-axis. 
The bars to the right indicate typical 2$\sigma $ uncertainties at levels of 5 and 10 \xbu.}
\label{fig6}
\end{figure}
%
\section{Discussion}
As mentioned in Sect.\,3.2, attempts to disentangle the spectral nature of either extended 
X-ray source in NGC\ 7714 remained inconclusive due to the low angular separation of the knots 
and the relatively low number of received counts. Keeping in mind the uncertainties discussed 
in Sects.\,3.1\&3.2 we shall assume that both knots are emitting thermal brems\-strahlung with 
a mean plasma temperature $\sim 0.5$ keV ($\sim 6\times 10^6$ K).
Then the X-ray luminosity of component E is $L_{\rm X}^{\rm E}{\rm \,\,(0.1-2.4\, keV)}\sim 8\times 10^{40}$ \ergsec,  
while for the nuclear source we shall adopt the upper estimate of 
$L_{\rm X}^{\rm N}{\rm \,\,(0.1-2.4\, keV)}\sim 2\times 10^{41}$ \ergsec\ (cf. Sect.\,3.2).
In Sects.\,4.1\&4.2 we shall investigate physical processes accounting for the morphological and energetic
X-ray properties of each emitting component. 
%
\subsection{Nuclear energy source}
From the surface brightness profile of NGC\ 7714 (Fig.\,2) we estimate 
the extinction corrected magnitude of the starburst nucleus to 13.5~mag. 
At the adopted distance of 40~Mpc to NGC\ 7714 the latter value translates to 
an absolute magnitude of $M_{\rm R} \approx -19.5$ mag.
%
\begin{table}
\caption[]{Properties of the X-ray components}
\begin{tabular}{lll}
\hline
Component & N &  E \\
\hline
Net counts               & 221$\pm$22                 &  112$\pm$16     \\
Count Rate (counts\,\,ksec$^{-1}$) & 4.85$\pm$0.46  & 2.48$\pm$0.36 \\
Softness Ratio$^{\alpha}$      & 1.1$\pm$0.2              &  1.6$\pm$0.5 \\
r$_{50}^{\beta}$     & 6\farcs 4$\pm$0\farcs 2   &  5\farcs6$\pm$0\farcs 2  \\
r$_{80}^c$           & 10\farcs 9$\pm$0\farcs 25 &  9\farcs 0$\pm$0\farcs 3 \\
f$_{\rm X}^d$                 & $\sim$11.5             &  $\sim 4$            \\
L$_{\rm X}^e$           & $\sim $2                     &  $\sim 0.8$ \\
\hline
\end{tabular}

\vspace*{1ex}
${\alpha}$: HRI {\it softness ratio} determined as the ratio of the net counts received 
in the HRI channels 1--5 and 6--11 (cf. Wilson et al. 1992).\\
${\beta}$,$c$: Radii enclosing 50\% and 80\% of the flux, as determined from the intensity
profiles (Fig.\,7).\\ 
$d$;$e$: Intrinsic flux and luminosity in the ROSAT energy band (0.1--2.4 keV) in units of
10$^{-13}$ \uflux\ and 10$^{41}$ \ergsec, respectively. Both quantities are derived on the 
assumption that the X-ray emission in either component can be approximated by a \emph{thbr} model
and shares the same spectral properties.
\end{table}
%
According to Bernl\"ohr (1993) the photometric properties of the nuclear region in NGC\ 7714 can be accounted 
for by an ongoing burst of star formation of age $\tau_{\rm burst} \sim 20$ Myr. Scaling the model predictions of 
Leitherer \& Heckman (1995, hereafter LH95) to the observed nuclear luminosity and assuming continuous 
star formation over $\tau _{\rm burst}$, we obtain an average star formation rate (SFR) of $\approx 2.2$ \msun\,yr$^{-1}$. 
Hereby, we adopt a Salpeter initial mass function ($\alpha =2.35$), solar metallicity and lower and upper 
stellar mass cutoffs of 1\,\msun\ and 100\,\msun, respectively.
We shall remark that the above model-dependent quantities are approximative only and represent a 
lower limit for the actual SFR in NGC\ 7714 as being derived by other authors from detailed spectrophotometric 
studies. For instance, from the \ha-luminosity of NGC\ 7714  Storchi-Bergman et al. (1994) deduced a SFR
of 17.2\,$(75/{\rm H_0})^2$ \msun\,yr$^{-1}$. Furthermore, Calzetti (1997) has inferred from UV-to-NIR synthesis studies
a SFR of 2.8$\div$12.2\,$(75/{\rm H_0})^2$ \msun\,yr$^{-1}$ and deduced a much lower starburst age of $\sim 10$ Myr. 

Scaling Figs.\,2\&6 of LH95 to the adopted SFR we obtain for the present evolutionary age of the nuclear starburst a
number $N_{\rm O*}\approx 7\times 10^4$ of O stars and a SN rate $\dot{SN} \approx 0.022$ yr$^{-1}$.
The corresponding production rate of \emph{mechanical} energy injected into the ISM by massive stars and SNe is 
$\dot{E}_{\rm m}\approx 8.8\times 10^{41}$ \ergsec (LH95, Fig.\,56) and the total \emph{mechanical} energy generated
since the onset of the starburst is $E_{\rm m}\approx 4.4\times 10^{56}$ erg (LH95, Fig.\,58).\\
The X-ray luminosity from O stars is $\sim N_{\rm O*}\cdot 10^{33}$\ \ergsec\ 
$\simeq 7\times 10^{37}$ \ergsec.
Given an average X-ray luminosity of $L_{\rm X}^{\rm SNR}\approx 2\times 10^{36}$ \ergsec\ over a time scale 
$\tau _{\rm SNR}\sim 2\times 10^4$ yr for a SN remnant (Cowie et al. 1981, Williams et al. 1997) 
we estimate the contribution of such sources to 
$\tau _{\rm SNR}\cdot \dot{SN} \cdot L_{\rm X}^{\rm SNR} \approx 9\times 10^{38}$ \ergsec\ or $<1$\% of $L_{\rm X}^{\rm N}$.
From the number ratio of $c_1\sim 2\times 10^{-3}\dots 10^{-3}$ of O-stars to 
High Mass X-ray Binaries (HMXBs; Fabbiano et al. 1982) 
we obtain an upper limit of 140 currently active HMXBs in the nuclear starburst region of NGC\ 7714. With a  
typical X-ray luminosity of a HMXB of $ \sim 10^{38}$ \ergsec\ this results in a maximum luminosity contribution 
of this type of objects of the order of $1.4\times 10^{40}$ \ergsec.
In summary, the direct contribution of X-ray emitting point sources produced by the nuclear starburst, such 
as O stars, SNRs and HMXBs amounts to $\sim 1.5\times 10^{40}$ \ergsec\ or $\la 10$\% of $L_{\rm X}^{\rm N}$. 
This percentage will not be greatly enhanced even when postulating the presence of a few 
\emph{super-Eddington} point-like sources with luminosities of $1.4 - 5\times 10^{39}$\ \ergsec\ as 
found in some nearby starburst galaxies (Dahlem et al. 1994, Vogler \& Pietsch 1997, Wang et al. 1995, 
Fabbiano et al. 1997). 

As a result, $\sim 90$\% of the X-ray luminosity confined to the starburst nucleus of NGC\ 7714 must be attributed to 
the emission by extended hot gas (few $\times 10^6$ K) being heated up by the injection of \emph{mechanical} 
energy from OB stars and SNe.
Equation\,3.2 of Nulsen et al. (1984) yields for the required mass of the hot plasma in the 
nuclear region of NGC\ 7714
\begin{equation}
\begin{small}
\frac{M_{\rm hot}^{\rm N}}{\rm M_{\odot}} \approx 1.73\times 10^5\, \left(\frac{r}{\scriptstyle 100 {\rm pc}}\right)^{\textstyle \frac{3}{2}}
\left(\frac{L_{\rm X}^{\rm N}}{\scriptstyle 10^{40}}\right)^{\textstyle \frac{1}{2}}\left(\frac{\Lambda (T)}{\scriptstyle 5\times 10^{-23}}\right)^{\textstyle -\frac{1}{2}}\,\eta^{\frac{1}{2}}
\end{small}
\end{equation}
where $r$ is the radius of the volume in pc, $\Lambda(T)$ the emission measure in 
$10^{-23}\,\,{\rm erg\,s^{-1}\,cm^3}$, and $\eta$ the volume filling factor.
Assuming for the emitting volume of component N a radius of 1 kpc and a mean plasma temperature 
$\sim 6\times 10^6$ K we obtain a nominal gas mass $M_{\rm hot}^{\rm N} < 3\times 10^7\times \sqrt{\eta}$ \msun.
Thus, the mass of the hot plasma required to explain the nuclear X-ray luminosity of NGC\ 7714 is comparable to 
the stellar mass formed in the burst. 
The radiative output of the hot gas ($\sim 2 \times 10^{41}$\ \ergsec) corresponds to roughly 20\% of the 
current \emph{mechanical} energy output provided by the nuclear stellar population. 
Thus, there is no fundamental energy problem for the starburst hypothesis. 
\subsection{Eastern X-ray component}
The absence of signatures of recent star formation at the stellar ring close to component 
E (Sect.\,1) excludes the possibility that its X-ray luminosity of $\sim 8\times 10^{40}$ \ergsec\ 
is due to a superposition of point sources such as HMXBs and SNRs.
Next we shall investigate whether the extended emission observed in region E is not intrinsic to 
NGC\ 7714 but due to the superposition of at least two foreground or background sources. 
As pointed out in Sect.\,2, the estimated detection limit of the CCD-frame (Fig.\,1) is 
$\sim 20.3$ R\,mag. This limit can effectively be pushed down by $\ga 1$ mag when applying the 
hb-method. At this level, ground-based images do not reveal any optical point sources 
coincident with component E. 
The only conspicuous feature seen in an archival HST-WFPC2 exposure of NGC\ 7714 
at a distance $\delta_{x,y}$=+16\farcs 7, +5\farcs 1 from the
nucleus is a pair of two faint point sources being separated by $\approx 0\farcs 2$. 
Their apparent R magnitude, estimated to 22.8 mag, is equivalent to that of a
M5V star ($M_{\rm R}=+10.5$; Johnson 1966) at a distance of $\sim 2.9$ kpc.
The X-ray luminosity of such late-type stellar sources amounts to $5\div 15\times 10^{27}$ 
\ergsec\ (Schmitt \& Snowden 1990, H\"unsch \& Schr\"oder 1996). 
At the assumed distance of 2.9 kpc, a source of this type would produce 
an X-ray flux $\sim 10^{-17}$ \uflux, much lower than the detection threshold of ROSAT.
Analogously we can excluse other candidates of foreground and background sources as
origins for component E. We have made detailed checks on the possibilities of 
the contamination by Pop\,II binaries, background clusters and diffuse background sources (Hasinger et al. 1993).

We, therefore, conclude that the extended X-ray flux detected at the location of component E be 
a superposition of point sources of known type to be extremely improbable.
Given that a physical intersection between the eastern \hi-bridge (Smith et al. 1997) 
and NGC\ 7714 is likely, we consider below
(\emph{i}) a starburst wind interaction with 
the low-density halo or with the dense \hi-bridge and (\emph{ii}) collisions of \hi\ clouds resulting 
from gas infall from the bridge including the interaction of the rotationally moving disk 
clouds with the material of the bridge.
%
\subsubsection{Starburst-driven outflow}
%
Theoretical and observational work in the past decade has advanced a picture where 
gas heated up by the central energy source is able to penetrate the ambient
\hi-layer and then to escape into the galactic halo with velocities of the order of $10^3$ km\,s$^{-1}$ 
(Chevalier \& Clegg 1985, Heckman et al. 1987, Norman \& Ikeuchi 1989 and references therein) 
where it produces an extended shock region. This mechanism could cause component E. 
The presence of wind of warm gas with a velocity $\sim 150$ km\,s$^{-1}$ with its axis 
pointing to the NE of the nuclear region is already suggested by optical spectroscopy 
(Ta\-ni\-gu\-chi et al. 1988). Although the spectra do not prove a large-scale outflow of hot gas 
towards the galactic halo they show, however, the gas phase in NGC\ 7714 to be kinematically perturbed.

Suchkov et al. (1994, hereafter SBHL94) have shown that in a nuclear starburst-wind 
the bulk of the soft (0.2--2.2 keV) X-ray photons will not be produced in the wind 
material itself but rather by its collisionally shocked interface with the ambient halo gas. 
By analogy to the model prediction of SBHL94, component E could be interpreted as the shell of a hot 
cavity expanding from the nuclear starburst, with the wind material itself being too hot ($\sim 10^8$ K) to 
make a measurable flux contribution to the ROSAT energy band. Read \& Ponman (1998) have presented a 
thorough discussion on unipolar outflows of hot gas seeming not to be unusual among colliding 
galaxies in a mid-stage of merging. An unipolar structure seems to readily evolve when 
injection occurs above the galaxy plane (Mac Low \& McCray 1988) or in the presence 
of a kinematically disturbed ISM.  

Models A1 and A2 of SBHL94 are computed assuming starburst properties 
similar to those we have adopted for NGC\ 7714, e.g. a SFR of 2 \msun yr$^{-1}$, Salpeter IMF and 
$\tau_{\rm burst}=17$~Myr and yield a number of 
predictions in accord with the present observations. 
For instance the luminosity of component E can be accommodated by 
a wind-halo interaction model when assuming a halo density $<\!5$ cm$^{-3}$.
However, in this model it is expected that the source will appear much more extended 
than the rather compact component E.

Alternatively, component E may be the result of the collision of a starburst-driven 
wind with the western boundary of the \hi-bridge discovered by Smith et al. (1997).
This would lead to the formation of a shocked interface at the radius where the ram 
pressure ($\rho _{\rm wind}\,u_{\rm wind}^2$) of the wind becomes equal to the ambient pressure of the \hi-gas. 
As discussed in Tenorio-Tagle \& Mu\~noz-Tu\~n\'on (1997), while the outer shock 
is going on sweeping up the ambient cold gas, the inner (reverse) shock will thermalize the wind 
to temperatures of $1.26\times 10^8\,\,u_3^2$ K where $u_3$ is the wind velocity in units of $3\times 10^3$ \kmsec. 
In this model, the adopted plasma temperature $\sim 6\times 10^6$ K corresponds to a post-shock temperature 
generated by a velocity of the order of $ 700$ km\,s$^{-1}$. Pressumably, such a process is going to produce 
a relatively compact shocked interface with size comparable to the diameter 
of the western tip of the \hi-bridge, yielding a better correspondence to the 
morphology of component E than the wind-halo interaction hypothesis.  

In summary, the morphological and energetic properties of component E may be understood 
in terms of the collision of a starburst-driven wind with the eastern \hi-bridge rather than collision 
with the tenuous halo-gas. 
\subsubsection{Cloud-cloud collisions in the cold gas}
A further hypothesis for the origin of the X-ray emission in region E invokes 
gas infall from the eastern \hi-bridge on the \hi-disk of NGC\ 7714. 
As shown by Tenorio-Tagle (1982) the energy-release caused by an \hi-cloud 
impinging on the disk can provoke a supersonic expansion of a hot 
($10^6$--$10^7$ K) ring-shaped cavity in the ambient cold medium. 
Smith et al. (1997) were not able to confirm a systematic motion of the 
massive ($1.5\times 10^9$ \msun) \hi-bridge towards NGC\ 7714. 
Nevertheless, they measured a projected velocity dispersion $>100$ km\,s$^{-1}$ 
within the \hi-bridge. 
Therefore, if the \hi-bridge is intersecting the disk of NGC\ 7714 high-speed
gas cloud collisions are likely to occur in the interaction zone.

If the infalling gas medium is composed of clouds with a radius $r_{\rm c}\sim 5$ pc and 
a volume filling factor $\eta =0.1$ then the mean free path $(2/3)\,(r_{\rm c}/\eta )$
of an interloping cloud amounts to $\sim 30$ pc, thus less than the disk thickness of NGC\ 7714. 
Given the low sound speed in such a cloud ($\sim\!0.7$ \kmsec; Spitzer 1978), a collision at a relative velocity 
$v_{\rm rel}\sim $ few 100 \kmsec\ would be characterized by a high Mach number, thus, give rise 
to a strong shock wave propagating within either cloud with a velocity $(4/3) (v_{\rm rel}/2)$ 
(Jog \& Solomon 1992). From the latter considerations, a strip of shocked material with size 
comparable to $r_{50}$ ($\sim 1$ kpc) can be formed in a time span of few Myr when $v_{\rm rel}\sim 200$ \kmsec. 
Substituting the estimated intrinsic X-ray luminosity of component E and such a radius in Eq.\,(1) we obtain 
for the mass of the hot interior $M_{\rm hot}^{\rm E} \sim 1.5\times 10^7\times \sqrt{\eta }$ \msun. 
From Eq.\,(3) of Nulsen et al. (1984) the corresponding gas cooling time is 
$\tau _{\rm cool}\sim 15\,\sqrt{\eta }$ Myr. 
An estimate of $\sim 3.8\times 10^{55}\times \sqrt{\eta}$ erg for the energy content of component E 
can be inferred assuming a constant radiation of $\dot{E}\equiv L_{\rm X}^{\rm E}\simeq 8\times 10^{40}$\ergsec\  
over $\tau _{\rm cool}$. This is equivalent to the kinetic energy of a cloud with mass 
$\sim 6\times 10^7$ \msun\ ($\sim 4$\% of the mass of the \hi-bridge) 
plunging onto the disk of NGC\ 7714 with a velocity $v_{\rm rel}$.\\
From the virial mass of $2\times 10^{10}$ \msun\ within a radius of 15\arcsec\ inferred for NGC\ 7714 
by Bernl\"ohr (1993) and assuming rotational motions one obtains at the distance 
of component E a circular velocity of $\sim 150$ km\,s$^{-1}$. Thus, the required collisional 
velocity $v_{\rm rel}$ is already comparable to the average circular gas velocity in the disk of NGC\ 7714.\\
As a result, the release of kinetic energy at region E can be understood in a twofold manner:
gas infall onto the disk by analogy to Galactic \emph{worms}
(Heiles 1979, Mirabel 1982) and/or collision of rotationally moving \hi-clouds of the disk of 
NGC\ 7714 with the western boundary of the \hi-bridge.
Clearly, a number of parameters must be specified to assess the energy output of these processes. Most 
important, the \hi-kinematics and the thermalization efficiency during such a collision remains uncertain. 
Nevertheless, we argue that in view of the energetics such a mechanism powered by cloud collisions only 
can readily account for the observed X-ray luminosity in component~E.
%
\subsubsection{On the fate of component E}
%
Reinfall of tidally ejected matter is likely to be a frequent occurrence in 
the merging of two gas-rich galaxies (Hibbard \& Mihos 1995)
and may well be accompanied by processes discussed 
in Sects.\,4.2.1\&4.2.2 generating X-ray emission.
However, it is difficult to analyze how such processes depend on 
the kinematical, dynamical and geometrical circumstances of 
the merging event.
While Read \& Ponman (1998) do not report the presence of any strong
extranuclear features they find the clear presence 
of low-surface-brightness extended X-ray emission in their sample.
They discuss the possibility that such X-ray halos seen in evolved mergers 
do not form entirely through the starburst-driven ejection of hot gas from the nucleus,
but also through infalling shock-heated tidal matter.

If component E is due to a systematic gas infall from the bridge 
one may speculate about the observability of this feature. 
Among the fifteen thoroughly investigated colliding systems
shown in Fig.\,5 only Arp\ 284 exhibits a 
high-surface-brightness extranuclear X-ray component.
Assuming that the formation of such features in the interaction region 
must commonly occur at some stage during the collision lasting $\sim 10^9$ yr,
then from a statistical point of view this feature cannot survive much
longer than $10^8$ yr.

This time-scale does not seem unreasonable:
as discussed above, the initial impact between two cold media with a velocity of 
a few $\times$ 100 \kmsec\ can readily give rise to the formation of a shocked 
interface. Subsequent infall of cold gas occurring with a velocity smaller 
than the internal sound speed in the hot cavity is not expected to add much to the
shock-induced luminosity. The period of visibility of such a shock-region 
(identified with region E of Arp\ 284)
will be of the order of its cooling time $\tau_{\rm cool}\la 10^8$ yr.
While mass-loading of the hot cavity enhances the cooling efficiency, differential 
rotation may smear the high-surface-brightness X-ray features (again on a time-scale $\sim 10^8$ yr),
thus contributing to the formation of a diffuse low-surface-brightness X-ray halo
similar to those found by Fricke \& Papaderos (1996) and Read \& Ponman (1998).
%
\section{Summary and conclusions}
Numerical simulations suggest that Arp\ 284 is the result of an off-center encounter between the 
spiral galaxies NGC\ 7714/15 (Smith et al. 1997). The less massive one, NGC\ 7715, underwent 
a burst of star formation $\sim 70$ Myr ago, presumably at the time of the closest passage between both systems.
While NGC\ 7715 is at present in a quiescent post-starburst phase, its interacting neighbour NGC\ 7714 
(Smith et al. 1997) being by a factor $\sim 3$ more massive harbours an ongoing nuclear 
starburst initiated $\la 20$ Myr ago (Bernl\"ohr 1993, Calzetti 1997). 

Morphological signatures of the interaction in Arp\ 284 are a conspicuous stellar ring in NGC\ 7714 
to the east of its nucleus and a faint ($\mu _{\rm R} \ga 23$ \sbu) stellar bridge probably 
connecting NGC\ 7715 and NGC\ 7714. 
Smith et al. (1997) have discovered a massive ($\sim 1.5\times 10^9$ \msun)
\hi-bridge parallel to the stellar one, likely being of tidal origin.
Both features intersect NGC\ 7714 roughly 20\arcsec\ eastwards of its nuclear region, near to the 
outer boundary of the stellar ring.\\
The surface brightness distribution of NGC\ 7714 shows two distinct luminosity parts 
in excess of its disk brightness. On small scales ($R^*\la 5$\arcsec) the light 
is being dominated by its compact high surface brightness 
($\mu _{\rm R}< 16$ \sbu) starburst nucleus. 
At intermediate intensity levels and out to a radial extent equivalent to two 
disk scale lengths the intensity distribution of NGC\ 7714 shows an initially 
flat and then sharply decreasing plateau.
This feature is partly due to the luminosity of the stellar ring mentioned above.

In agreement to results from \emph{Einstein}, ROSAT~PSPC observations  
show that the entire emission in Arp\ 284 is confined to NGC\ 7714.
Fits to the ROSAT~PSPC spectrum imply a thermal emission origin approximated 
best by a thermal bremsstrahlung model with $k$T$\sim $0.4$\div$0.6 keV ($4.6\div 7\times 10^6$ K).  
We deduce an intrinsic 0.1--2.4 keV luminosity of NGC\ 7714 ranging between 2 and $\sim 4.4\times 10^{41}$ \ergsec, 
comparable to the luminosity reported for other interacting/merging starburst systems. 

ROSAT~HRI maps reveal that the soft X-ray luminosity of NGC\ 7714 is being contributed by 
two distinct X-ray emitting sources both of which were found to be extended. 
The more luminous one (N), for which we estimate an X-ray luminosity 
$\la 2\times 10^{41}$ \ergsec\ is coincident with the starburst nucleus
of NGC\ 7714.
Its luminosity can be accounted for by the superposition of point- and diffuse 
X--ray emitting sources expected to form in a starburst. 
Point sources evolving in a continuous nuclear starburst
with an age $\sim 20$ Myr and average SFR $\sim $ 2 ${\rm M_{\odot }}$\ yr$^{-1}$
are estimated to provide only a minor fraction ($\sim 10$\%) of the nuclear X--ray luminosity in NGC\ 7714 
while the bulk of the soft X-ray emission is being produced by hot ISM residing in 
a volume $\la 2$ kpc in diameter.

The second X--ray source (component E), located rough\-ly 20\arcsec\ eastwards of 
the starburst nucleus is of intriguing nature. It lacks any optical counterpart and the
probability for confusion with an assembly of X-ray point sources unrelated 
to Arp\ 284 is low.
We estimate the intrinsic X-ray luminosity of this component to 
$\sim 8\times 10^{40}$ \ergsec\ and favour a model based on \emph{in situ} 
shock--heated gas. With respect to the combined evidence of VLA- and ROSAT~HRI maps 
different interpretations seem possible. 
One plausible scenario invokes the collision of a starburst--driven wind with 
the western boundary of the massive \hi-bridge. In this case a relatively  
compact shocked layer with a temperature of few 10$^6$ K may evolve.
Alternatively, gas--infall from the eastern \hi-bridge 
onto the disk of NGC\ 7714 with a speed comparable to the intrinsic disk velocity 
dispersion can produce a collisional interface powering component E.
Collisional breaking of the disk-rotation in NGC\ 7714 by \hi-clouds contributes 
an additional energy source.
We argue that the X-ray spot is a transient phenomenon and may eventually evolve into 
an extended low-surface-brightness X-ray halo as commonly observed in merging systems.
\begin{acknowledgements}
This work was supported by the 
Deut\-sche Agen\-tur f\"ur Raum\-fahrt\-ange\-legen\-heiten (DARA) grant 50~OR~9407~6.
We wish to thank an anonymous referee for useful comments which have helped to 
clarify various aspects of this work.
This research has made use of the NASA/IPAC extragalactic database (NED) which is operated by the 
Jet Propulsion Laboratory, Caltech, under contract with the National Aeronautics and Space Administration.
\end{acknowledgements}


\begin{thebibliography}{}
   \bibitem[1966]{arp66} {Arp, H.C. 1966, Atlas of Peculiar Galaxies (Pasadena: Caltech)}
   \bibitem[1992]{ba92}{Barnes, J.E., Hernquist, L. 1992, ARA\&S 30, 705} 
   \bibitem[1996]{ba96}{Barnes, J.E., Hernquist, L. 1996, ApJ 471, 115}
   \bibitem[1993]{bern1} {Bernl\"ohr, K., 1993, A\&A 268, 25}
   \bibitem[1995]{biko95} {Bischoff, K., Kollatschny, W., Pietsch, W. 1996, 
in {\sl R\"ontgenstrahlung from the Universe}, eds. H.U.\ Zimmermann, J.E.\ Tr\"umper, H.\ Yorke, 423}
   \bibitem[1990]{bus90} {Bushouse, H.A., Werner, M.W. 1990, ApJ 359, 72}
   \bibitem[1997]{calz97}{Calzetti, D. 1997, AJ 113, 162}
   \bibitem[1985]{chev85} {Chevalier, R.A., Clegg, A.W. 1985, Nature 317, 44}
   \bibitem[1981]{cow81} {Cowie, L.L., McKee, C.F., Ostriker, J.P. 1981, ApJ 247, 908}
   \bibitem[199*]{dahl94} {Dahlem, M., Hartner, G.D., Junkes, D. 1994, ApJ 432, 598}
   \bibitem[1992]{david92} {David, L.P., Jones, C., Forman, W. 1992, ApJ 388, 82}
   \bibitem[1993]{david93} {David, L.P., Harnden, F.R.,Jr., 
Kearns, K.E., Zombeck, M.V. 1993, The ROSAT High Resolution Imager (HRI) 
(Technical Rep., US ROSAT Science Data Center/SAO)}
   \bibitem[1989]{dev89}{Devereux, N.A., Eales, S.A. 1989, ApJ 340, 708}
   \bibitem[1990]{dl90} {Dickey, J.M. \& Lockman, F.J. 1990, ARA\&A 28, 215}
   \bibitem[1994]{mir94}{Duc, P.-A., Mirabel, I.F. 1994 A\&A 289, 83}
   \bibitem[1982]{fab82} {Fabbiano, G., Feigelson, E., Zamorani, G. 1982, ApJ 256, 397}
 \bibitem[1992]{fab92} {Fabbiano, G., Kim, D.-W., Trinchieri, G. 1992, ApJS 80, 531}
 \bibitem[1997]{fab97} {Fabbiano, G., Schweizer, F., Mackie, G. 1997, ApJ 478, 542 }
   \bibitem[1989]{fk89}{Fricke, K.J., Kollatschny, W. 1989, Proc. of IAU Symp. No. 134, 425}
   \bibitem[1995]{fri96} {Fricke, K.J., Papaderos, P. 1996, in {\sl R\"ontgenstrahlung from the Universe}, 
eds. H.U.\ Zimmermann, J.E.\ Tr\"umper, H.\ Yorke, 377}
   \bibitem[1995]{delgado95} {Gonz\'alez-Delgado, R.M., P\'erez, E., D\'iaz, \'A.I., 
Garc\'ia-Vargas, M.L., Terlevich, E., Vilchez, J.M. 1995, ApJ 439, 604}
   \bibitem[1996]{gu96} {G\"udel, M., K\"urster, M. 1996, in {\sl R\"ontgenstrahlung from the Universe}, eds. 
H.U.\ Zimmermann, J.E.\ Tr\"umper, H.\ Yorke, 37}
 \bibitem[1991]{has91}{Hasinger, G., Tr\"umper, J., Schmidt, M. 1991, A\&A 246, L2} 
 \bibitem[1993]{has93}{Hasinger, G., Burg., R., Giacconi, R., Hartner, G., Schmidt, M., 
Tr\"umper, J., Zamorani, G. 1993, A\&A 275, 1} 
 \bibitem[1987]{heck87} {Heckman, T.M., Armus, L., Miley, G.K. 1987, AJ 93, 276}
 \bibitem[1990]{heck96} {Heckman, T.M., Dahlem, M, Eales, S.A., Fabbiano, G., Weaver, K. 1996 ApJ 457, 616}
 \bibitem[1979]{hei79} {Heiles, C.E. 1979, IAU Symp. 184, ed. W.B.\ Burton, 301}
 \bibitem[1992]{he92} {Hernquist, L. 1992, ApJ 400, 460}
  \bibitem[1989]{hb89}{Hewitt, A., Burbidge, G. 1989, ApJS 69, 1}
   \bibitem[1995]{hm95} {Hibbard, J.E., Mihos, J.C. 1995, AJ 110, 140} 
   \bibitem[1995]{hib95} {Hibbard, J.E., van Gorkom, J.H. 1996, AJ 111, 655} 
   \bibitem[1988]{hig88}{Higdon, J.L. 1988, ApJ 326, 146}
   \bibitem[1996]{hie96}{H\"unsch, M., Schr\"oder, K.-P. 1996, A\&A 309, L51}
   \bibitem[1992]{jog92} {Jog., C.J., Solomon, P.M. 1992, ApJ 387, 152}
   \bibitem[1966]{jo96} {Johnson 1966, AR\&AA 4, 193}
   \bibitem[1996]{kol96} {Kollatschny, W., Kowatsch, P., Fricke, K.J. 1996, in {\sl R\"ontgenstrahlung from the Universe}, eds. H.U.\ Zimmermann, J.E.\ Tr\"umper, H.\ Yorke, 383}
   \bibitem[1995]{lh95} {Leitherer, K., Heckman, T.M. 1995, ApJS 96, 9}
   \bibitem[1978]{lt78} {Lynds, R., Toomre, A., 1976, ApJ 209, L382}
   \bibitem[1988]{mac}{Mac Low, M.-M., McCray, R. 1988, ApJ 324, 776}
   \bibitem[1996]{mihe96} {Mihos, J.C., Hernquist, L. 1996, ApJ 464, 641}
   \bibitem[1982]{mir82} {Mirabel, I.F. 1982, ApJ 256, 112}
   \bibitem[1994]{mo94} {Morse, J.A. 1994, PASP 106, 675}
   \bibitem[1989]{nor89} {Norman, C., Ikeuchi, S. 1989, ApJ 345, 372}
   \bibitem[1989]{n96} {Norman, C.A., Bowen, D.V., Heckman, T., Blades, C., Danly, L. 1996, ApJ 472, 73}     
   \bibitem[1984]{nul984} {Nulsen, P.E.J., Steward, G.C., Fabian, A.C. 1984, MNRAS 208, 185}
   \bibitem[1996]{p98} {Papaderos, P. \& Fricke, K.J. 1998, in preparation}
   \bibitem[1987]{pfef87} {Pfefermann E., Briel U.G., Hippmann H., et al., 1987, Proc. SPIE 733, 519}
   \bibitem[1976]{rs76} {Raymond, J.C., Smith, B.W. 1977, ApJS 35, 419}
    \bibitem[1995]{read95} {Read A.M., Ponman T.J., Wolstencroft R.D. 1995, MNRAS 277, 397}
    \bibitem[1995]{read95} {Read A.M., Ponman T.J., Strickland, D.K. 1997, MNRAS 286, 626}
   \bibitem[1998]{rp98}{Read, A.M., Ponman, T.J. 1998, MNRAS 297, 143}
  \bibitem[1997]{fer97}{Sanders, D.B., Mirabel, I.F. 1996, ARA\&A 34, 749}
   \bibitem[1990]{sch90}{Schmitt, J.H.M.M., Snowden, S.L. 1990, ApJ 361, 207}
   \bibitem[1982]{sch82}{Schweizer, F. 1982, ApJ 252, 455}
   \bibitem[1991]{smi91} {Smith, B.J. 1991, ApJ 378, 39}
   \bibitem[1992]{smi92} {Smith, B.J., Wallin, J.F. 1992, ApJ 393, 544}
   \bibitem[1991]{smi97} {Smith, B.J., Struck, C., Pogge, R.W. 1997, ApJ 483, 754}
   \bibitem[1992]{so92}{Solomon, P.M., Downes, D., Radford, S.J.E. 1992, ApJ 387, L55}
   \bibitem[1978]{spi78} {Spitzer, L. 1978, Physical Processes in the Interstellar Medium (New York: John Wiley)}
   \bibitem[1994]{storch94}{Storchi-Bergmann, T., Calzetti, D., Kinney, A.L. 1994, ApJ 429, 572}
   \bibitem[1994]{su94} {Suchkov, A.A., Balsara, D.S., Heckman, T.M., Leitherer, C. 1994, ApJ 430, 511}
   \bibitem[1990]{su90}{Sulentic, J.W. 1990, IAU Colloquium 124, 291}
   \bibitem[1988]{tan88} {Taniguchi, Y., Kawara, K., Nishida, M., Tamura, S., Nishida, M.T. 1988, AJ 95, 1378}
   \bibitem[1996]{tt96} {Telles, E., Terlevich, R. 1997 MNRAS 286, 183}
    \bibitem[1997]{ten82} {Tenorio-Tagle, G. 1982, A\&A 88, 61}
    \bibitem[1997]{ten97} {Tenorio-Tagle, G., Mu\~noz-Tu\~n\'on, C. 1997, ApJ 478, 134}
    \bibitem[1977]{tom77}{Toomre, A., Toomre, J. 1972, ApJ 178, 623}
   \bibitem[1983]{trumper83} {Tr\"umper J., 1983, Adv.\ Space\ Res. 2, 241}
   \bibitem[1988]{tully88}{Tully, R.B. 1988, Neraby Galaxies Catalog, Cambridge University Press, Cambridge}
   \bibitem[1994]{vel94} {Veilleux, S., Cecil, G., Bland-Hawthorn, J., Tully, R.B., Filippenko, A.V., Sargent, W.L.W. 1994, ApJ 433, 48}
   \bibitem[1997]{vog92} {Voges, W., Gruber, R., Paul, J. et al. 1992, in {\sl Proc.\ European ISY conference}, 
eds. T.D. Guyenne, J.J. Hunt, 223}
   \bibitem[1997]{vog97} {Vogler, A., Pietsch, W. 1997, A\&A 319, 459}
    \bibitem[1997]{wang95}{Wang, Q.D., Walterbos, R.A.M., Steakley, M.F., Norman, C.A., Braun, R. 1995, ApJ 439, 176}
    \bibitem[1997]{wang97}{Wang J., Heckman, T.M., Weaver, K.A., Armus, L., 1997 ApJ 474, 659}
   \bibitem[1981]{weed81} {Weedman, D., Feldman, F.R., Balzano, V.A., Ramsey, L.W., Sramek, R.A., Wu, C.-C. 1981, ApJ 248, 105}
   \bibitem[1995]{wichu95}{Williams, R.M., Chu, Y.-H., Dickel, J.R., Beyer, R., Petre, R., Smith, R.C., Milne, D.K. 1997, ApJ 480, 618}
   \bibitem[1992]{wils92}{Wilson, A.S., Elvis, M., Lawrence, A., Bland-Hawthorn, J. 1992, ApJ 391, L75}
  \bibitem[1994]{zim94}{Zimmermann, H.-U. et al. 1994, EXSAS User Guide, 4th edition} 
\end{thebibliography}
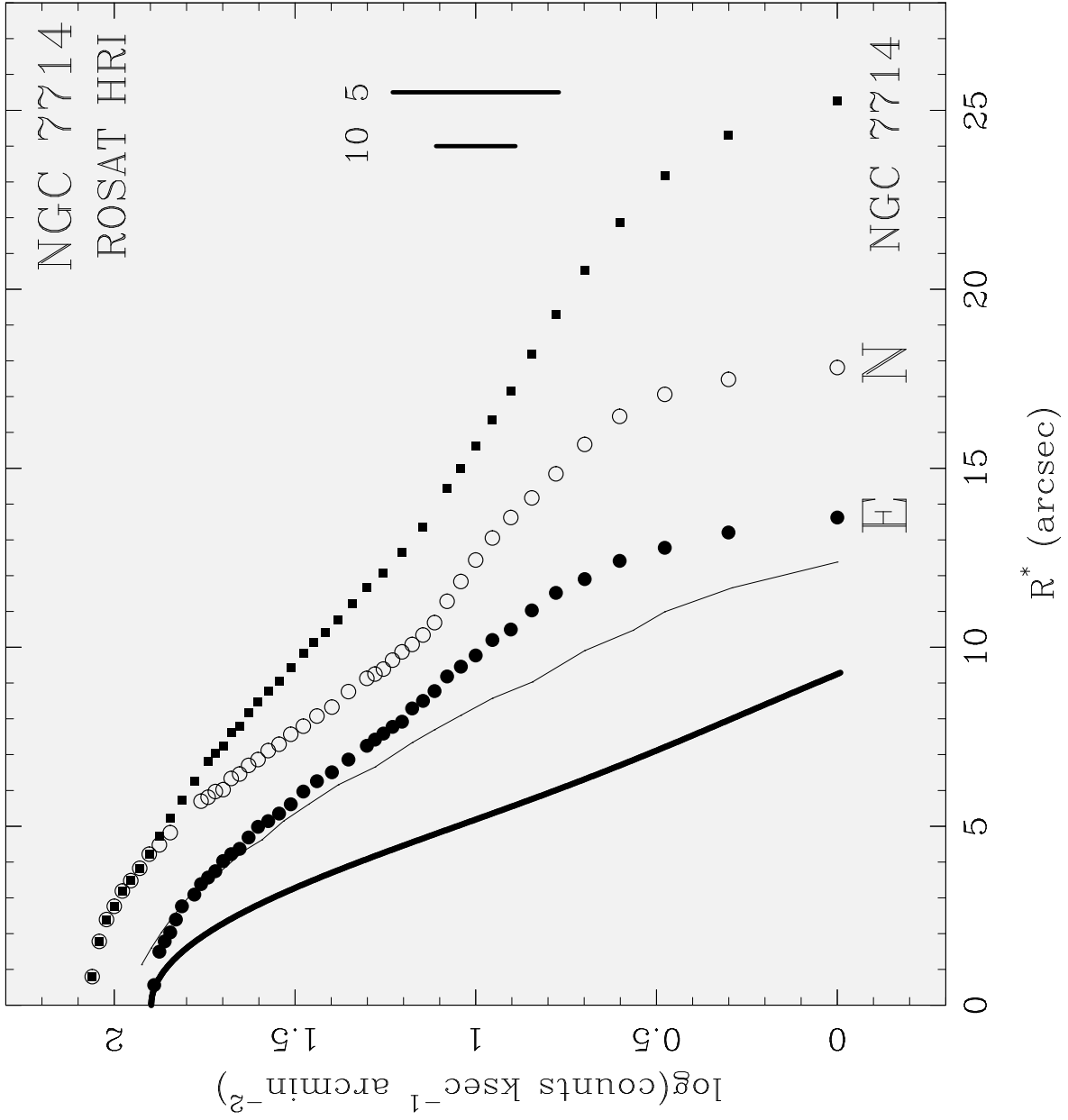
\end{document}